Snowmass 2013 White Paper

# Ultra-Low Index and Rad-Hard Total Internal Reflection Claddings for High Numerical Aperture Scintillating, Waveshifting, Cerenkov, or Clear Optical Fibers, Capillaries, and Plates

C. Sanzeni, G. Jennings, D.R. Winn*
Fairfield University
Fairfield, CT 06880 USA

*Corresponding Author
winn@fairfield.edu
1+203.984.3993

*Abstract:* Novel low index (n≤1.3)/high light trapping cladding films consisting of nanoporous alumina (sapphire), as formed by controlled anodization of aluminum, are described. These films are mechanically hard, intrinsically very rad-hard, and have an index of refraction n sufficiently small to triple the light capture of waveshifting or scintillating fibers and transparent plastic, glass or quartz core fibers. The low indices enable light-piping of water Cerenkov light. Applications are in Energy, Intensity, and Cosmic/Astroparticle detectors.

*Introduction:* Optical fiber and light-transmitting Total Internal Reflection (TIR) technology is frequently used in High Energy, Astroparticle, and Nuclear Physics research. Uses include organic scintillating (SciFi) and wavelength shifting (WLS) fibers, quartz Cerenkov fibers, and clear fibers, used to transport analog signals. They are used in calorimetry, rpreradiators and tracking, often in high magnetic fields, and others, including in telescope spectroscopy image planes. Generally, except for quartz core/doped quartz claddingsfibers, they have low radiation hardness, and a modest numerical aperture (N.A.), especially doped quartz cladded fibers(NA~0.2). We propose to extend radiation hardness to ultra-low index (high NA>0.8) claddings for superior light capture, increased by as much as a factor of 3 in transmitted light fractions. This cladding technology is easily applied to scintillating, WLS or Cerenkov plates, fibers or capillaries.

*Low Index Rad-Hard Fiber Claddings:* An extensive review of fibers used in HEP experimentation is given by H.Leutz[1], CERN. Typical plastic fiber performance is shown in the figures below.

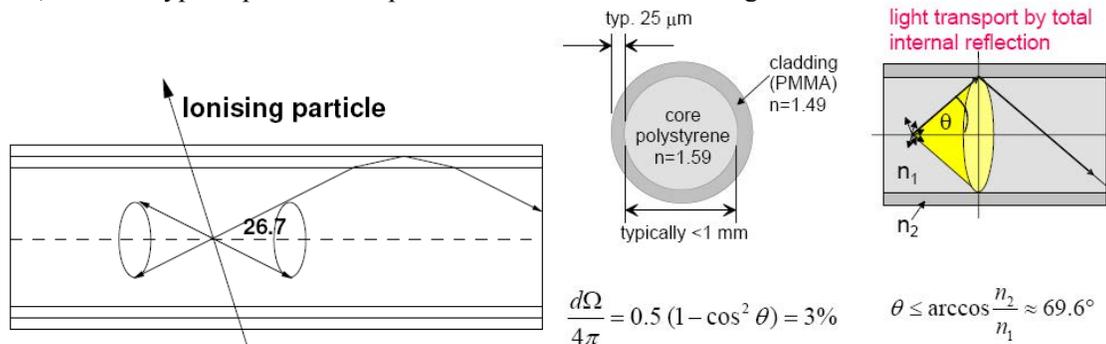

**Fig. 1:** Cartoons of light trapping in a typical plastic WLS fiber[1]. The angle θ is the trapped cone angle.

The fraction of light piped to each end of a fiber is shown in the cartoon of Fig.1. The fraction f piped is given by the cladding $n_2$ and the core $n_1$, the critical angle α and the Numerical Aperture NA by:

$$f = \tfrac{1}{2} (1 - n_2/n_1). \qquad (1)$$
$$\sin \alpha = n_2/n_1 \qquad (2)$$
$$NA = (n_1^2 - n_2^2)^{1/2} = \sin \theta \qquad (3)$$

Typical fractions per end are f~3% for commercial plastic fibers. If the cladding could be made to have an index of 1.3, the fraction trapped would rise to ~9%, *tripling the trapped light fraction.* A cladding index of 1.4 would double the fraction of trapped light, to ~6%, as compared to the example with a polystyrene core and a PMMA cladding as shown above. The number of reflections in terms of the diameter core d length of fiber L critical angle $\alpha$ by: $N = \cot \alpha \, L/d$. For small diameter fibers, the imperfections in the walls dominate. The reflection loss length $L_R$ over which the injected intensity is reduced by 1/e from reflection imperfection losses 0<q<1 (i.e q is the effective the reflectivity back into the trapped region) is given by: $L_R \sim 1.5/(1-q) \, (n_1/NA) \, d$. For many scintillators and claddings this will dominate over scintillation light self-absorption for small diameter fibers. In general, L, the 1/e total attenuation length is given by: $L^{-1} = L_A^{-1} + L_R^{-1} + L_s^{-1}$, where $L_s$, $L_A$ are the scattering and absorption lengths. The cladding must be >4 wavelengths thick.

**Novel Low Index of Refraction Radiation-Hard Hardcoating Technology for Optical Material Total Internal Reflection Claddings.**

To obtain a very low index cladding film, we have developed a novel method to anodize completely metal films applied to insulators, so that no metal remains, except for an arbitrarily small connection line at an edge of the work piece. The controlled anodization of metals under proper conditions results in a hard metal-oxide film with an ordered nanoporosity, up to ~70% porous in hexagonal close packing, with the pore diameters well-less than 100 nm < $\lambda_{light}$, and the pore lengths up to ~100 times longer than the diameters (i.e. porous film thickness ~µm's ~several optical $\lambda$). The resulting index n is dominated by the index of the air in the pores, and thus is low (see below).

The formation of arrays of high aspect ratio nanopores during the anodic oxidation of ultra-pure aluminium is a well-known phenomenon. The occurrence of a near-periodic pore arrangement in porous anodic alumina was first reported by Masuda et al. in the early 1990's (see refs [2] and loc. cite). The areal density, size, and regularity of these pores is controllable over a remarkably wide range with the electrolyte chemistry, temperature, current density, and the electric field during the anodization. Recent work is shown in Figures 3. Films of Al ~ 3 µm thick resulted in uniform porous structures, the pore diameter being in the range of 10–30 nm and their density of the order of $8\text{-}9 \times 10^{10}$ pores cm$^{-2}$. Pore widening with phosphoric acid[3] can easily increase the open area to above 70%.

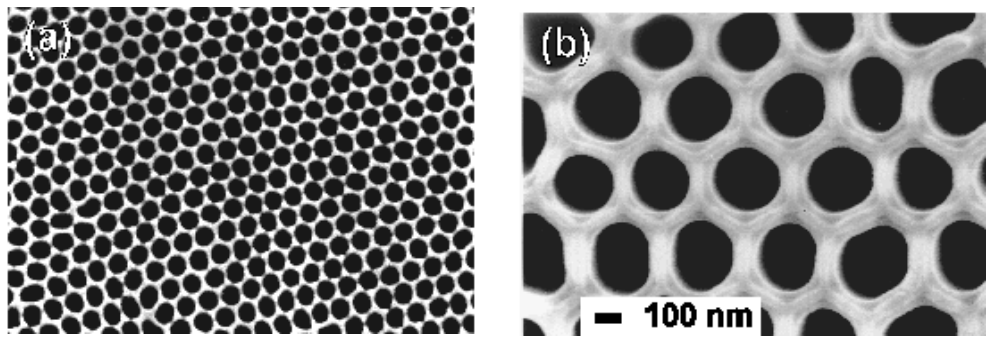

**Figure 3a, b:** Anodic alumina nanoporous films[4]. Left – 30 nm pores. Right ~130 nm pores.

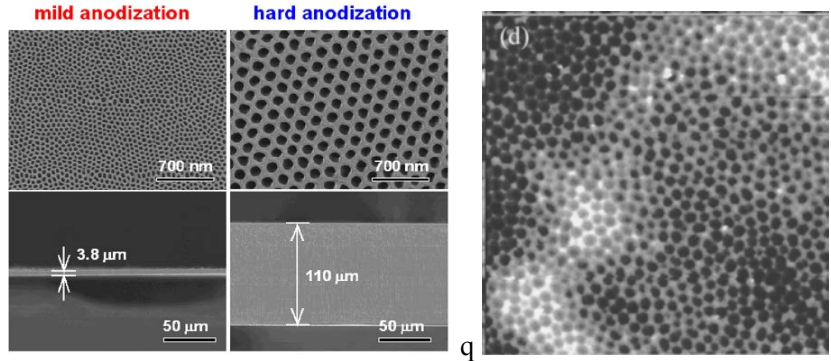

**Figure 3c(Left):** Ultra-thin porous anodic alumina films with 2 anodization protocols, with self-ordered cylindrical vertical pores on a silicon substrate[5] Note the film thicknesses. The pores sizes are as compared with the 700nm fiducials. Either film can be pore-widened with post anodization isotropic etch. **Fig. 3d(Right):** A 2 μm x 2 μm square SEM field of anodic alumina. The pores are ~50 nm, < optical wavelengths.

Because the pores are far smaller than visible wavelengths, the resultant heterogenous films can be considered uniform for the optical properties of visible light, and have an index of refraction for heterogeneous media given by several possible formulae[6] which bracket the possible extremes - (a) Parallel: $n \sim fn_p + (1-f)n_o$ ; (b) Series: $1/n \sim f/n_p + (1-f)/n_o$ ; or (c) Drude: $n^2 \sim fn_p^2 + (1-f)n_o^2$ where f is the porous fraction of the film, $n_p$ is the pore index, equal to that of air (i.e. $n_p = 1$) unless filled, and $n_o$ is the oxide matrix index. For the form of Aluminum Oxide here (boehmite, γ-alumina), for example, $n_o \sim 1.75$, and so for f = 65% porosity, easily achieved, the index of the film is bracketed by n in the ranges ~ 1.26/1.18/1.3, respectively as calculated for the 3 formulae above, low enough to capture significantly more light in fibers with such a cladding. NOTE: For n<1.33, such a material could pipe light from a water core n=1.33 *(if the pores were sealed with an obliquely applied film, as commonly used in anodic products, or coated with an ALD hydrophobic film),* a significant additional tool for large neutrino & proton decay detectors. As discussed above, this low cladding increases the trapping fraction significantly and enables far better detection even with the thinnest fibers[7]. *At cladding n=1.3 with a core index of 1.6, the NA=0.93, and 9% of the light is transmitted, a factor of 3 over the best plastic fibers. For quartz cores (n=1.54), 8% of the light is transmitted. For a water-core light guide, with a cladding of 1.26, 2.7% of light is transmitted.*

Normally a metal film-on-insulator, or any metal, when immersed in an anodization bath and connected to an anode power supply anodizes to form an insulator (metal oxide) layer at the surfaces in contact with the electrolyte. The anodized layer, even if porous, then prevents further anodization from occurring, unless new metal is exposed to the electrolyte, *and* a conducting connection is supplied to the remaining metal. Even in the case of porous anodization, a boundary layer of metal must be left; as the metal anodizes, the ability to conduct current is reduced to near zero, leaving metal surrounded by anodized metal. Using the standard anodizing technique, it would be impossible to anodize a metal film on an insulator, such as a quartz fiber or plate completely - i.e. leaving no unanodized metal. For example, an aluminum film on an insulator would anodize until either islands of metal remained, or a continuous film of metal remained on the insulator. Even if a through-insulator connection to the back side of the metal were provided, isolated islands of unanodized metal would result.

However, we have demonstrated a simple method to anodize a metal film on an insulator or thin plate of aluminum completely. This method can also be used to control the currents on the sample surface by limiting the area of aluminum presented to the electrolyte in the anodization process. In principle, the technique can be implemented in several ways. The method we have demonstrated in our lab is shown schematically in Fig.4 below:

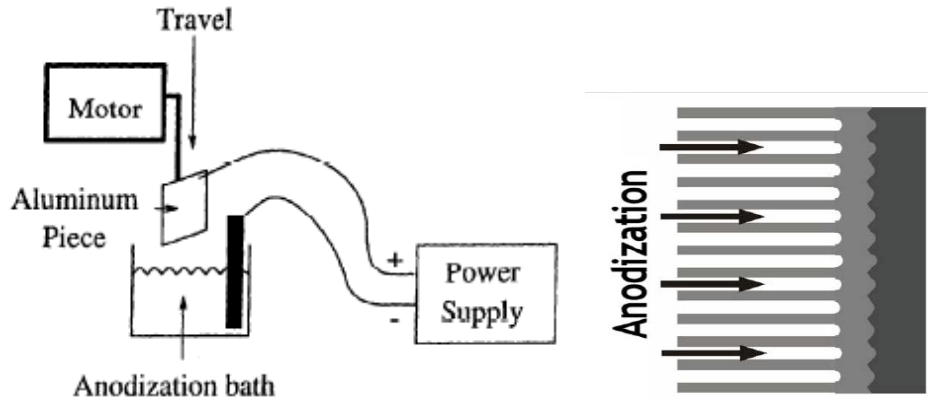

**Figure 4:** Schematic view of the edge anodization technique where aluminum pieces are controllably submerged into the anodization bath, in our experiment by a feed-back driven stepper motor. The cathode connection to the electrolyte (anodization bath) is represented by the thick electrode, usually a lead or carbon sheet.

The thin aluminum piece (i.e foils) or aluminum film on an insulator, such as a transparent plastic scintillator or WLS fiber core, is mounted on a motorized stage which allows the sample to be lowered into the anodization bath at a controlled velocity. The key idea is that the section of the piece just above the anodization bath acts as the anode terminal connection, an edge connection equal at all parts of the cross-section through the thickness of the aluminum piece. As the piece is "adiabatically" lowered into the bath, in effect the anode terminal is presented equally to all parts of the unanodized metal in the bath, unlike the situation if the anodization starts on a fully immersed film, where metal not in direct contact with the electrolyte becomes insulated by the barrier oxide formed on the metal surrounding it. Only the thin stripe of aluminum which is just below the bath surface is unanodized, but connected to the same potential. The entire metal film is thus able to be anodized as the metal is immersed methodically into the electrolyte. *We call this technique Edge Anodization or EA.*

The key technical and scientific challenge is to provide control of the insertion motion to anodize the film with uniform quality. This motion can be controlled with feedback on the anodization current or current density. There are several constraints on the sample velocity. If the velocity is set too high, then the cell current/voltage (I/V) characteristic will be dominated by the current drawn as the virgin aluminum surface develops its oxide barrier. In the limiting case of high velocity, it is the same as anodizing the entire surface at once. If the velocity is too low, then the quality of the aluminum surface and the electrolyte/ aluminum interface will determine the quality of the anodized film. Extremely low velocities and unclean surfaces could lead to unanodized areas of the aluminum surface.

*Anodized aluminum and the formation of alumina and sapphire-like films:* The anodization of aluminum is a large subject, beyond the scope of this brief review. In general, aluminum anodizes to form a porous film of amorphous (non-crystalline) phases alumina (boehmite, γ-alumina - $Al_2O_3$) on the aluminum. The film is composed of essentially the same materials-base as sapphire and alumina ceramics. These materials are among the most intrinsically hard known, and are also transparent. Remarkably, the size of the resulting porous structure is essentially unchanged from the parent aluminum. Aluminum is not removed - the incorporation of oxygen during anodization produces a molecular structure smaller than the original aluminum, thereby providing the driving mechanism to create pores in the anodized aluminum. Micro/Nano/porous alumina ($Al_2O_3$) thin sheets or films exhibit a highly anisotropic and uniform pore structure consisting of channels perpendicular to the film surface, whose size and spatial distribution can be controlled by changes in the anodization process parameters[8,9,10]. The channel position is quasi-regular, on a quasi-regular hexagonal matrix. The areal density, size and regularity of these pores is controllable over a remarkably wide range with the electrolyte chemistry, temperature, current density, and the electric field during the anodization [11,12,13,14,15]. The pores typically range in size from ~5 nm-200 nm in size, and vary in areal open pore density from <1% of the alumina surface up to ~68-70%. The pores terminate in a non-porous thin boundary layer of alumina at the aluminum alumina interface, the thickness of which is also determined by the anodization conditions. Figure 5a,b shows a schematic of the anodization process and an SEM of an anodized aluminum film produced by this process(see refs in figs). These porous structures have wide application in technology (filters for example), and are the basis of colored anodized aluminum by incorporation of dyes into the pores.

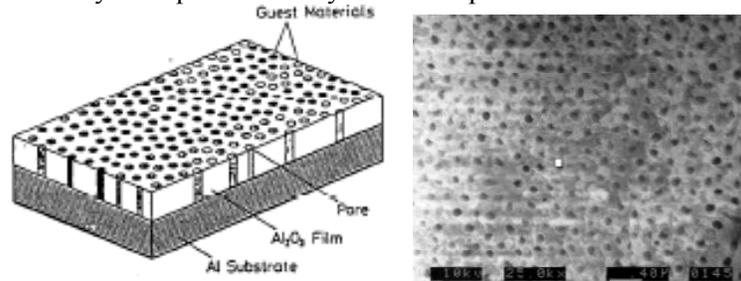

**Figure 5a(L):** Schematic of an anodized aluminum film produced by a total immersion method.[16] **Figure 5b:** SEM of a porous alumina film, produced by the edge anodization technique (EA). The nanoscale of the resulting hard film is evident. The fiducial is 400 nm in size and the pores are ~50-100 nm in size, smaller than relevant optical wavelengths. [17]

The physics behind the Edge Anodization technique (EA) is shown in Fig.6. As the aluminum piece is lowered into the bath, the resistance may be modeled as a parallel network of resistors. There are three distinct regions in the bath that characterize the resistor network. The first region is near the bath surface. This region will have the lowest resistance because new aluminum surface is constantly being introduced into the electrolyte. In fact, the only appreciable resistance presented by this region is the resistance of the native oxide present on the aluminum surface.

The second region lies below the bath surface where the initial solid alumina (Al2O3) barrier layer has formed. This region will present the highest resistance in the cell for the thin layer anodization because this is where the low conductance oxide is thickest. The third region is the porous alumina region which will have a resistance intermediate to the top two regions because the barrier oxide at the bottom of the pores is thinner than the oxide in the second region.

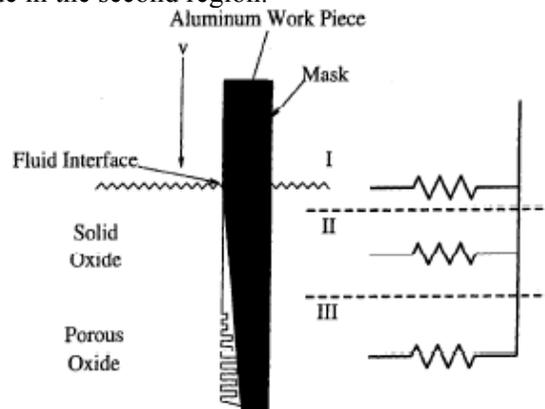

**Figure 6:** Schematic illustration of the anodization cell resistance. Note the three distinct regions on the sample representing the three stages of anodization as discussed in the text.

As the sample is lowered into the bath the size of the first region will remain unchanged. The velocity of the sample entry will determine the size of the other 2 regions. Higher velocities will allow the solid barrier region to be larger since we have determined that a finite time is required for the intial oxide barrier to reach its maximum thickness. The higher the initial velocity of the sample, the larger this region will become because more of the sample will be introduced during the period of time required for the barrier to reach maximum thickness. Conversely, lower velocities will result in a smaller solid barrier region.

To first order, the size of the porous region will depend on the overall size of the area A to be anodized. Large area samples will inevitably have a porous region develop near the leading edge of the sample first introduced into the bath. The size of this region is also dependent on the sample velocity to second order, since the size of this region is the size of the entire submerged portion of the sample, less the first and second regions. That is, for long times t:

$$A_{porous} = A_{submerged} - A_{interface} - A_{barrier} \quad (1)$$

or, by substituting the size of the other regions

$$A_{porous} = (v_{tw}-v)t_{req}w - A_{barrier} , \quad (2)$$

where v is the velocity, w is the sample width at the electrolyte surface, and $t_{req}$ is the time required for the initial oxide barrier to reach its maximum thickness.

We have demonstrated this EA technique to obtain complete anodization of aluminum on glass, with sputtered and with evaporated aluminum films, varying from about 0.5-4 μm thick, roughly following the schematic of Fig. 5,6. A stepper motor was programmed with step size and step period to insert the slide into an anodization bath, starting from the bottom, and a programmable power supply was programmed with a current profile.

Figure 7a shows a photograph of an early test of an aluminized glass slide after anodization, and back illuminated. There are 4 regions from top to bottom, corresponding to 3 different anodization regions. The 3 regions starting from the bottom were created by 3 different insertion velocities (stepsize/step periods) and cell voltages.

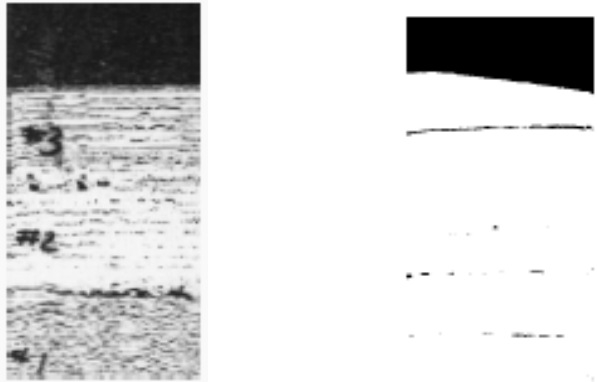

**Fig.7a (left):** Back-illuminated photograph of an aluminized and then edge anodized glass slide in early tests of edge anodization technique (EA), in 3 regions starting from the bottom. The 4th top region shows the opaque unanodized aluminum film used as the edge anode contact, as in Fig.2. The region second from the bottom (labeled with hand drawn " #2 ") demonstrates nearly complete anodization but shows that the stepper is not being run smoothly enough. Regions 1 and 3 were anodized with rates and currents which left larger and more numerous islands of aluminum unconverted to alumina. The wavy aluminum lines are due mainly to the unstable creep of the meniscus (or wetting line) as the work piece is inserted into the electrolyte anodization bath.
**Fig.7b (right):** Photograph of a back-illuminated (resting on a light box) aluminized glass slide in tests of complete edge anodization (ET), in 6 regions with different current density and insertion rates, starting from the bottom. The top region shows the opaque unanodized aluminum film used as the edge anode contact, as in Fig.2. Nearly all of the aluminum below it was converted to transparent alumina. At 4 of the boundaries between regions where the stepper changes speed and the current is changes, some aluminum lines are left. A hardened steel scribe did not scratch these surfaces, whereas a carbide scribe scratched.

The opaque top region shows the unanodized aluminum which was not inserted into the bath. The region second from the bottom shows evidence of nearly complete anodization of the aluminum. The banded regions at the bottom and third from the bottom reflect the stepper step size and anodization and insertion speed not optimal for complete anodization. Fig.7b shows the results of refined programming of the EA rig, showing nearly complete anodization in 6 regions of different EA conditions - the boundaries of some regions are nearly invisible.The transparent hard films appeared to be highly scratch resistant with metal tools, and required a carbide scribe. At some larger thicknesses, coloured interference fringes were evident. Angular reflection and transmission data taken with a HeNe laser and a Si detector indicated n<1.4.

Figure 7c shows a photograph of a free-standing alumina film created by edge anodization of a standard aluminum foil about 10 μm thick obtained from a local supermarket. It was anodized completely over a length of about 2.5 cm from the far end. To demonstrate its flexibility and clarity, it was bent or curled to rest conformally against a curled dollar bill and flash photographed. George Washington's face is clearly seen through the anodized foil end. The index was measured to be n=1.37.

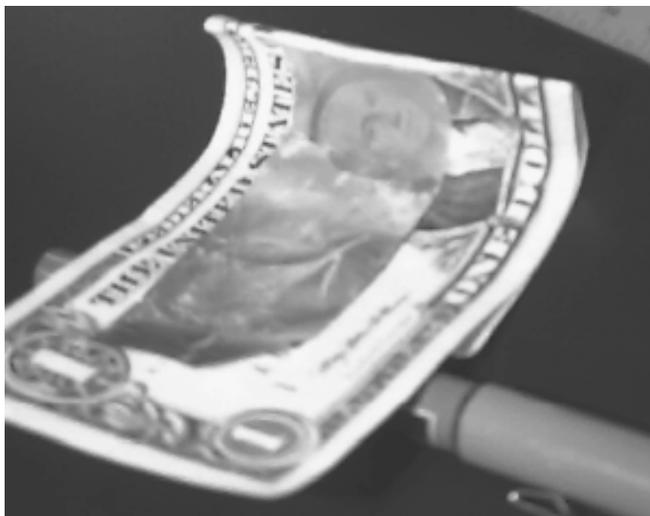

**Figure 7c:** Photograph of a George Washington's head on a dollar bill through a piece of free-standing alumina film, created from a strip of off-the-shelf Alcoa aluminum foil about 4 cm wide and ~10 cm long. The clear anodized film area is about ~ 2.5 cm long x ~3 cm wide starting at the edge of George Washington's face, covering his whole face and neck. The unanodized aluminum strip attached to it blocks an area ~ 5 cm x 3 cm starting at the right shoulder of the president and nicking the oval surrounding the "1" in the lower left of the picture. A pen is included for scale.

Such an immersion technique (EA) to form transparent nanoporous alumina films can be made to work on Al films deposited on fibers or plates provided: (i) sufficiently large aluminization "bell jar" is obtained to aluminize the work piece- such bell jars are common for glazing at least in sizes of 4' x 8', or tubular for fiber pulling; and (ii) sufficiently large anodization bath is obtained in which to edge immerse the work piece.

A possible disadvantage of the immersion technique is that the pores formed at the leading edge of the work-piece are guaranteed to be deeper and/or wider than those formed near the solution interface because they are exposed to the low pH electrolyte longer, and etch. The extent of this non-uniformity is not known and needs study. We predict that the anodizing of the pores shuts off once the aluminum is removed, and predict further removal of the alumina by the chemistry of the bath would not be a serious problem for work pieces ~1m or less. However, if the pores at the start begin to become too large and then merge during the process, the piece would periodically be withdrawn after a set distance of immersion and the pores protected from the anodization bath by a spin-on polymer (as used in VLSI processing) which could then be stripped off after full processing. The piece would then be re-immersed to the edge of the protected region (i.e. back to the bare aluminum film) and anodization continued. For a fiber work-piece, the lower end of the fiber could simply be guided to exit the bath after a sufficient anodizing time. A more complicated remedy for large plates would be a form-fitting elastic-lined flap/slot at the bottom of the tank, allowing entrance to a much diluted anodizing bath below the anodizing bath. We emphasize that a post-anodizing isotropic etch (phosphoric acid or similar) can widen the pores so that porosity exceeding 70% is possible.

*Application of Edge Anodization(EA) of Aluminum Films to Form Transparent Ultra-Low Index Cladding Hard Coatings on Scintillating/Cerenkov Plates:*
The application of the EA to transparent hard coatings on flat optical substrates is straightforward. Aluminum films of a few microns thickness are applied industrially to large area materials in sheets, films and highly complex shapes, most commonly using e-beam evaporators, and has been successfully scaled to continuous line production (eg mylar party balloons etc). Aluminum has a low melting temperature compared transparent with hard coatings, and adheres strongly to many clean plastic surfaces, including lexan and lucite and other polycarboates and acrylics. The aluminum is then converted by the EA to alumina/alumina-like materials. The alumina film produced by EA is essentially an amorphous sapphire or ceramic, one of the hardest materials. The Hardness of these materials varies between 10-20 GPa[18] On hard substrates, the Knoop Microhardness of $Al_2O_3$ films is quoted[19] typically as ~1,000 $kg/mm^2$.

*Application of Edge Anodization(EA) of Aluminum to Form Cylindrical Transparent Ultra-Low Index Cladding Hard Coatings on Scintillating, Cerenkov or WLS Fibers or Capillaries:*
Coating thin films of Al and many other materials on fibers, or on the inside or outside of capillaries is standard practice (flow-through MOCVD, ALD, or flood MBE, and various forms of PVD, often with rotating susceptors). The growth of parallel pores on a cylindrical surface by EA is a possible issue. There are 2 points:
  (a) The pores do follow the electric field, which in the fiber/capillary case will be radial – this is ok – if the pores are slightly further apart at the outside over the 3+ micron thick film, it makes little difference in the index.
  (b) The thickness of the aluminum film applied to the fibers is, say, ≥3 micron thick. If the fiber or capillary cores are minimum of 150 μm, then the E-field varies by ~2% or less over the thickness to be anodized, and locally the E-field is essentially flat to a few%.

*Adhesion and Surface finish:* Adhesion of the alumina film may be an issue. On glass or quartz, this did no seem to be an issue (Figs 7a,b). The pores are fine enough that scratches to the surface of the pores is negligible, on the scale of optical wavelengths.

*Z-Axis Connection:* An aside - filling the high aspect ratio pores with metal or other conductor forms a "z-axis" connector, an anisotropic conducting film or sheet which has very low resistance through the film (z-direction) and a high resistance in the x-y plane of the film. Thus 2 planar VLSI circuits can be connected by sandwiching such a film between 2 silicon substrates, *without top/bottom registration, if the pores are smaller and more dense than the circuit features*.

**Summary:** Transparent anodic films of the boehmite form of alumina, formed by the Edge Anodization technique, have an index of refraction dominated by the index of the pores formed therein, which if filled with air may have indices between n~1.25-1.4. In thicknesses >5λ (few μm for visible) these films function as claddings for high numerical aperture total internal reflection light transmission.